\documentclass[%
 reprint,
 amsmath,amssymb,
 aps,
 prd,
floatfix,
showkeys
]{revtex4-1}

\usepackage[USenglish]{babel}
\usepackage{slashed}
\usepackage{graphicx}
\usepackage{dcolumn}
\usepackage{bm}
\usepackage{amssymb}
\usepackage{amsmath}
\usepackage{graphicx}
\usepackage{epsfig}
\usepackage{rotating}

\usepackage[utf8]{inputenc} 
\usepackage{amsmath} 
\usepackage{array}
\usepackage{float}
\usepackage{color}
\usepackage[usenames,dvipsnames]{xcolor}
\usepackage{epstopdf}
\usepackage{natbib}
\usepackage{graphicx}
\usepackage{lineno}
\usepackage{slashed}
\usepackage{color}
\usepackage{hyperref}
\usepackage{float}
\usepackage{mathtools,slashed}

\newcommand{\be}{\begin{equation}}
\newcommand{\ee}{\end{equation}}
\newcommand{\bea}{\begin{eqnarray}}
\newcommand{\eea}{\end{eqnarray}}
\newcommand{\beas}{\begin{eqnarray*}}
\newcommand{\eeas}{\end{eqnarray*}}

\newcommand{\bse}{\begin{subequations}}
\newcommand{\ese}{\end{subequations}}

\begin{document}

\title{\bf Tumor and microcalcification characterization using Entropy, Fractal Dimension and intensity values statistical analysis in a mammography
}
\author{C.H.~Zepeda~Fern\'andez$^{1,2}$}
\email{Corresponding author, email:hzepeda@fcfm.buap.mx}
\author{M.G.~V\'azquez~Dom\'inguez$^{2}$}
\author{E.~Moreno~Barbosa$^2$}
\author{B. De Celis Alonso$^2$}
\author{M.~Rodr\'iguez~Cahuantzi$^2$}

\address{
$^1$Cátedra CONACyT, 03940, CdMx Mexico\\
$^2$Facultad de Ciencias F\'isico Matem\'aticas, Benem\'erita Universidad Aut\'onoma de Puebla, Av. San Claudio y 18 Sur, Ciudad Universitaria 72570, Puebla, Mexico\\
}

\begin{abstract}
\begin{center}
\bf \large Abstract
\end{center}
Digital analysis of mammographic images (MI) is a complementary tool to clinical evaluation, commonly used to identify tumors and/or microcalcifications in mammograms. Recent mammographic equipment, can automatically classify them using this methodology. The difficulty in finding and classifying such areas, arise from different factors such as: image acquisition methodology, excess of brightness, similar physiological and radiological properties of tissues, etc. In this work it is proposed that numerical computations of fractal dimension, entropy and pixels intensity data analysis are tools that could be used to automatically segment and distinguish malignant  tumors and/or microcalcifications (T/M) in a  MI from the rest of the breast (background). The study consisted in transform the MI into a 2D histrogram (MIH), to be able to segment the MIH in two areas: background and confirmed T/MC. To make the distinction, it was calculated the fractal dimension and entropy values in both areas. As a result, it was found the correlation between them. From each MIH, it was able to give the exact coordinates of the T/MC as the diagnosis provides: for all MIH, the highest intensity value was located in the T/MC-area, resulting that the fractal dimension had a higher value than the background, while the entropy value was lower in T/MC-area than in the background. To complete this study, it was performed a data analysis with the set of pixel intensity values in each area. This allowed to have a second technique to distinguish between the T/MC-area from the background. Resulting that, the intensity values located in the T/MC-area are  greater than $3\sigma$ from the mean value of the background distribution. Finally, it is shown a third technique to visually highlight the T/MC-area, which consisted in raising the intensity values to a power. With this three mathematical techniques, it is possible to characterize and distinguish a T/MC from a MI.
\end{abstract}


\maketitle

\section{Introduction}\label{1}
According to the \textit{World Cancer Research Fund},  breast 
cancer ranks second of all cancers based on its prevalence (lung cancer being the first) \cite{WCRF}, it is also the cancer that most 
commonly occurs in women. Due to this incidence, there is a great variety  of clinical studies developed and used to diagnose it.  Mammografic images (MI) are the gold standard in the clinic nowadays and are used to find anomalous masses, which can either be malignant or benign tumors. With this technique, it is possible to observe tumors,  microcalcifications, macrocalcifications, and fatty tissue. The image of a tumor appears in general lines as an amorphous mass, while a microcalcification appears as a dot and usually  is not a manifestation of cancer  unless there are many of them present in a structured pattern \cite{microcal}. In some cases, it could be difficult to distinguish it and make a diagnosis, due to fatty tissue, image acquisition methodology, excess of brightness, artifacts, etc.  Therefore, it is necessary to use other techniques to determine the malignant and/or benignant tumor and/or microcalcification (T/MC) location. The use of artificial intelligence (AI) analysis is  of high interest,  as it  can complement and make clinical diagnosis more precise~\cite{AI,AI2,MC-AI,AI3}. There are several techniques based in mathematical treatments to characterize the area of the tumor, one of them is to calculate the entropy value~\cite{entropy_1, entropy_2}. One of the definitions of entropy is the \textit{Shannon entropy} (S) from information theory~\cite{entropy}, which is given by the equation:

\begin{equation}\label{eqentropy}
    S=\displaystyle -\sum_{i=1}^{n}p_ilog_2p_i
\end{equation}

where $p_i$ is the $i$-value (probability) in an intensity map of the $i$-pixel (which it is normalized) and $n$ is the total pixels contained in the image. This definition can be applied to a section of the total image. S can be viewed as an uncertainty measure,  then, an image with a great number of pixels with the same intensity value, the S value will be lower compare it with a less uniform distribution. Through these features, the image can be considered as information carries.\\
The Fractal Dimension (FD) has been studied to  understand properties of  images~\cite{colorimage, physicspattern}. It has been applied in different topics to characterize different phenomena~\cite{naturalphenomena, plantdevelopment}. To consider that mammograms are digital images, they can be analyzed with topological properties. Therefore, it is possible to associate a FD to these images or a given region in them. These studies have been conducted for several years for tumors and microcalcifications~\cite{1989,1998}. As it was mentioned above, image processing has been very helpful in the case of tumors, for example, those that originate in glial cells, which replace conventional techniques of Magnetic Resonance and use FD to identify most malignant grades as FD increase~\cite{processing}. The box counting method is one of the most method used to obtain the FD, which has been improved by differential box counting applied to intensity map  images~\cite{DiffBox}. As an example, the FD has been used in other topics in medicine, for example, to detect branch retinal vein occlusion~\cite{branchretinal}. The use of FD is to find patterns in images or a section of them, this patterns are cancer risk indicators, then, it seeks to prevent it~\cite{prevent,AI3}. For this reason, this technique is very used, however, it is possible to analyze the FD in function of the brightness (pixel intensity) of the image viewed as a 2D  histogram or intensity map and then characterize the image through its texture~\cite{fractal}. This FD is calculated with:

\begin{equation}\label{fraceq}
    \displaystyle FD=2-\frac{log[A(\epsilon,i_\epsilon)]}{log[\epsilon]}
\end{equation}

where $A(\epsilon,i_\epsilon)$ is an area (not geometrical) of the image where the FD is calculated (geometrical area), defined as
\begin{multline}\label{area}
    \displaystyle A(\epsilon,i_\epsilon)=\sum_{x,y}\epsilon^2+
    \sum_{x,y}\epsilon(|i_\epsilon(x,y)-i_\epsilon(x,y+1)|+\\
    |i_\epsilon(x,y)-i_\epsilon(x+1,y)|)
\end{multline}
this sum is calculated over the location of each pixel $(x, y)$ of area $\epsilon\times\epsilon$ and considering its intensity value ($i_\epsilon$). Do not confuse the value of the region where the image to analyze is chosen with the value of  $A(\epsilon,i_\epsilon)$. The first one represents the size of the image (or a section) using the Euclidean metric, while $A(\epsilon,i_\epsilon)$ depends of the intensity and the size of the pixel of the image (or a section).
The counting boxes techniques do not consider the intensity value pixel information,  which, this quantity is essential for the purpose of this work. Then we used Eq.~\ref{fraceq} to calculate the FD.\\\\ 
 MI are obtained by the interaction of X-rays with breast,   which is matter. Due to the nature of this electromagnetic waves, they go through differently depending of the material of the breast. The image obtained shows different values in an intensity map which can be analyzed statistically (see Section ~\ref{map}), therefore, in this work is aimed at showing that it was possible to characterize T/MC-area and distinguish it from the rest of the breast (background). This would be achieved by calculating  the values of the FD and S using Eqs.~\ref{eqentropy}~and~\ref{fraceq}, respectively, to T/MC-area and comparing it with the values in the  background area (b-area) of the MI. 
Also, this work shows an intensity data analysis. It consists of characterizing the intensity values distribution in b-area to exclude it  for certain $sigmas~(\sigma)$  and thus keeping the T/MC-area. Finally, because the intensity values were normalized, it was possible to highlight the intensity values associated with the T/MC, by raising all values from the MI to some power, for contrast enhancement and thus highlight the values in T/MC.  There are shown three techniques to differentiate the T/MC-area from the MI.

 \section{Data selesction ans method}\label{2}
 
 \subsection{Data selesction}
 Five microcalcification cases were selected from patients of the  "Hospital de la Mujer" in Puebla. 54 MI of malignant tumors  were obtained from  \textit{PEIPA, the Pilot European Image Processing Archive: The mini-MIAS database of mammograms}~\cite{tumor_mammo}. Both sets of MI were complemented with their respective clinical diagnoses.
 
 \subsection{Image transformation method}\label{map}
  \subsubsection{MI 2D histogram}\label{2dh}
  In order to develop an algorithm to study the values of FD and S and in T/MC and b areas, the study began by transforming the MI from a PNG format into a intensity map using the CERN ROOT software version 6.23/01~\cite{root}, based on C++. The reference class used was  \textit{TASImage}~\cite{tasimage}, it allows to transform the original image to a several image format manipulation such as scaling, tiling, merging, etc., in a grid format, i.e., each pixel of the MI is assigned a normalized intensity values, in such a way that the intensity pixel information of the MI can be accesed. Then, it is possible to assign a coordinate system (no units) for the location of each pixel  and a palette defines the normalized intensity value of each one. The intensity map was created by the \textit{SetBinContent Class Reference}~\cite{setbin}, which it is a 2D histogram. In conclusion, it is obtained a 2D histogram from the MI (MIH). Thus, with this transformation, each pixel is well determined by knowing its location and its intensity value. Each transformation was conformed by 97,604 pixels, with a pixel size of $0.007\times0.007$.  As an example, in Figure~\ref{sigmapmt} it is shown the a MC MI and its respective MIH. The range of $x$ and $y$ axis can be freely chosen, for this work, we chose (-1,1) for both axis.
    \subsubsection{MI: intensity array}
    The set of intensity values can be stored in a square matrix, where each element has a one-to-one correspondence with the MIH. Each entry in the matrix corresponds to the position of a pixel of MI. This is the fundamental part of this work, because in this way data analysis can be done at the intensity values. As it is shown in Section~\ref{T-analysis}, the maximum intentity value element of the matrix can be compared with the coordinates of the central tumor location in a MI.\\
    In conclusion, we obtained:
    \begin{itemize}
    \item A visual representation (Subsection~\ref{2dh}) of the MI for qualitative treatment.
    \item An intensity values array of the MI for quantitative treatment.
    \end{itemize}
    \subsubsection{Tumor and microcalcification area selection}
\begin{figure}[htbp]
\begin{center}
\includegraphics[width=0.25\textwidth]{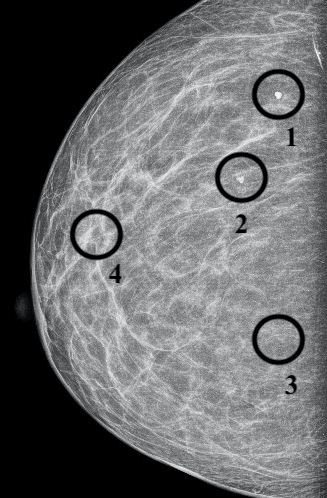}
\includegraphics[width=0.45\textwidth]{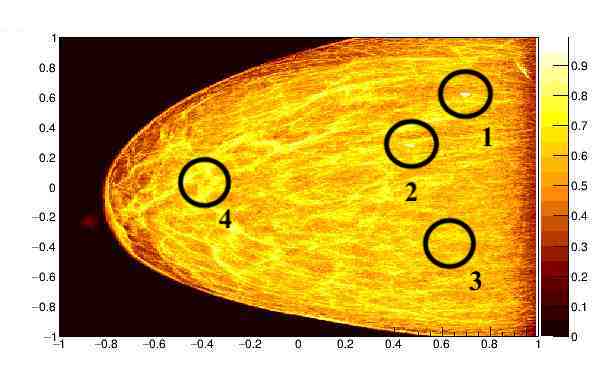}
\end{center}
\caption{ One of the mammographic images used in this study. (up) Mammographic image of microcalcification type  and (down) its transformation into a 3D phase-space. Top banner (1) MC, (2) C and (3) and (4) t. Bottom banner, same image with same tissues in phase-space.}
\label{sigmapmt}
\end{figure}

\begin{figure}[htbp]
\begin{center}
\includegraphics[width=0.3\textwidth]{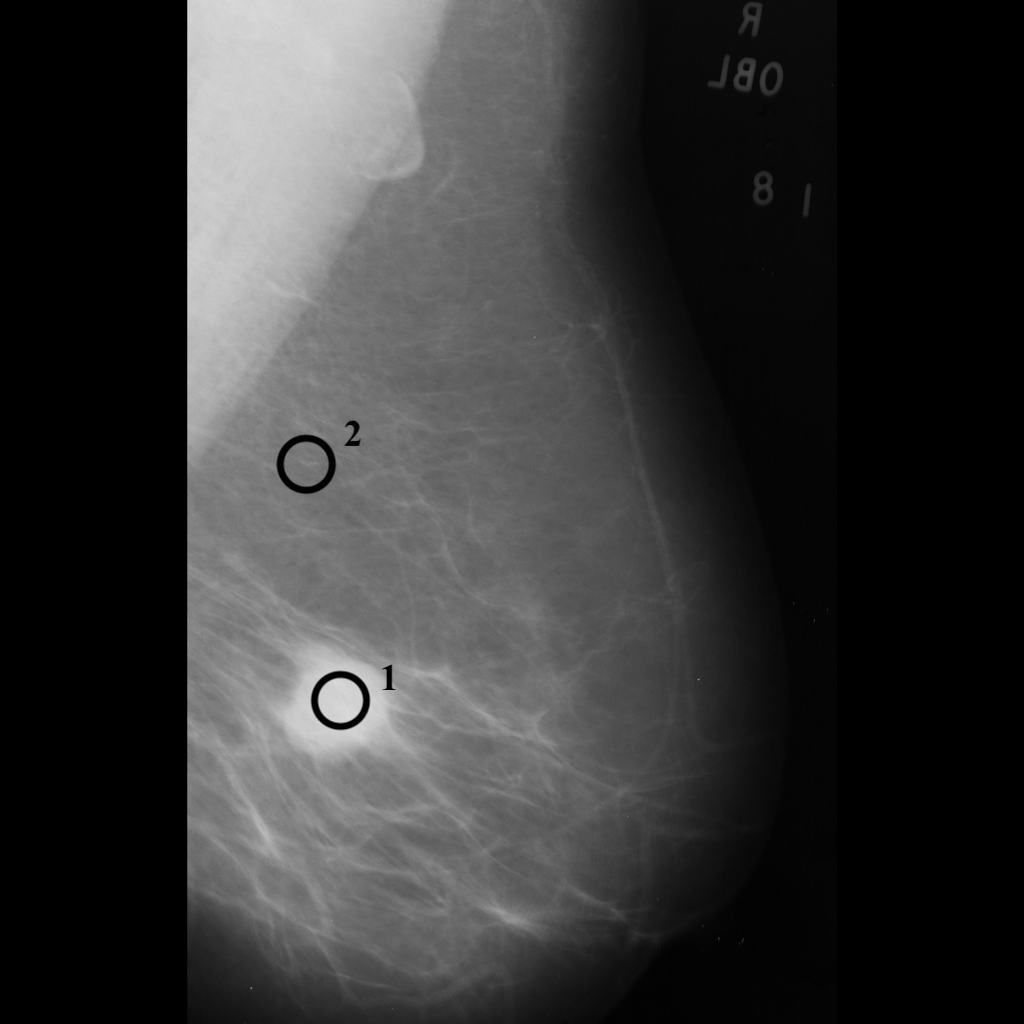}
\includegraphics[width=0.5\textwidth]{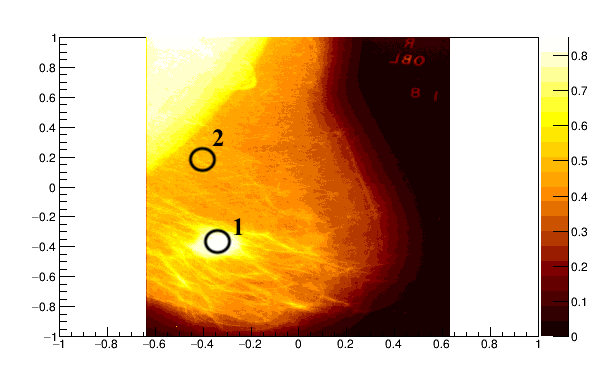}
\end{center}
\caption{The location of the (1) T and (2) t areas for (up) the original image and (down) its transformation.}
\label{areas2}
\end{figure}

    It was possible to enclose  in a ROI:  microcalsification (MC) as well as the whole malignant tumor tissue (T). This was done manually by an operator (no automatic segmentation was performed). The position and size of the ROIs matched the areas indicated by physicians in the clinical diagnostic. There were defined the background tissue (b) as the regions of the breast in which MC and T were not localized (b-area). For the case of the MC-Mammograms, it was possible to distinguished a third area, which it was  called MC candidate (C). In Figure~\ref{sigmapmt} these three areas  (MC, C and b) are presented, also it is shown a fourth b-area arbitrarily selected to compare the average intensities with the first b-area, as a result, they turned out to be consistent, then, the selection of any b-area is irrelevant. Same analysis was made also for T-mammograms, in this case,  a C-area was not considered, because it was not possible distinguish.\\
    For each MC-mammogram, the same ROI size to enclose the MC, C and b was chosen. For the T-mammograms, the region to enclose T was based on the clinical diagnosis and the same size for their respective b-area. In Figure~\ref{areas2} are shown the T and b areas for one of the T-mammograms. The areas showed in  Figures~\ref{sigmapmt} and \ref{areas2} are just a representation and not the actual ROI size  for the calculation. It should be noted that this technique is vulnerable to the so called mammographic artifacts, which can alter the analysis. For example, the letters found in Figure~\ref{areas2} were brighter than the breast  region. However, it was possible to isolate these areas to keep the breast region and then find the coordinates of the maximum intensity value. Once having the areas, the  FD and S values were calculated though the Equations~\ref{eqentropy} and \ref{fraceq}, respectively. Finally, with the set of intensity values in T/MC and b area, it was possible to obtain their intensity distributions and compare them to know the separation of their means for certain sigmas value ($\sigma$). As the intensities were normalized, it was possible to  raise the intensity values to certain powers in order to highlight the T/MC-area. These three quantities were used to characterize and distinguish T/MC and b areas as it is described below.

\section{Analysis and results}\label{3}
As a first and important result, from the MIH, it was possible to distinguish between the T/MC-area from the background, because the highest intensity value was located in T/MC-area. 

\subsection{T-mammograms analysis}\label{T-analysis}
\begin{figure}[htbp]
\begin{center}
\includegraphics[width=0.5\textwidth]{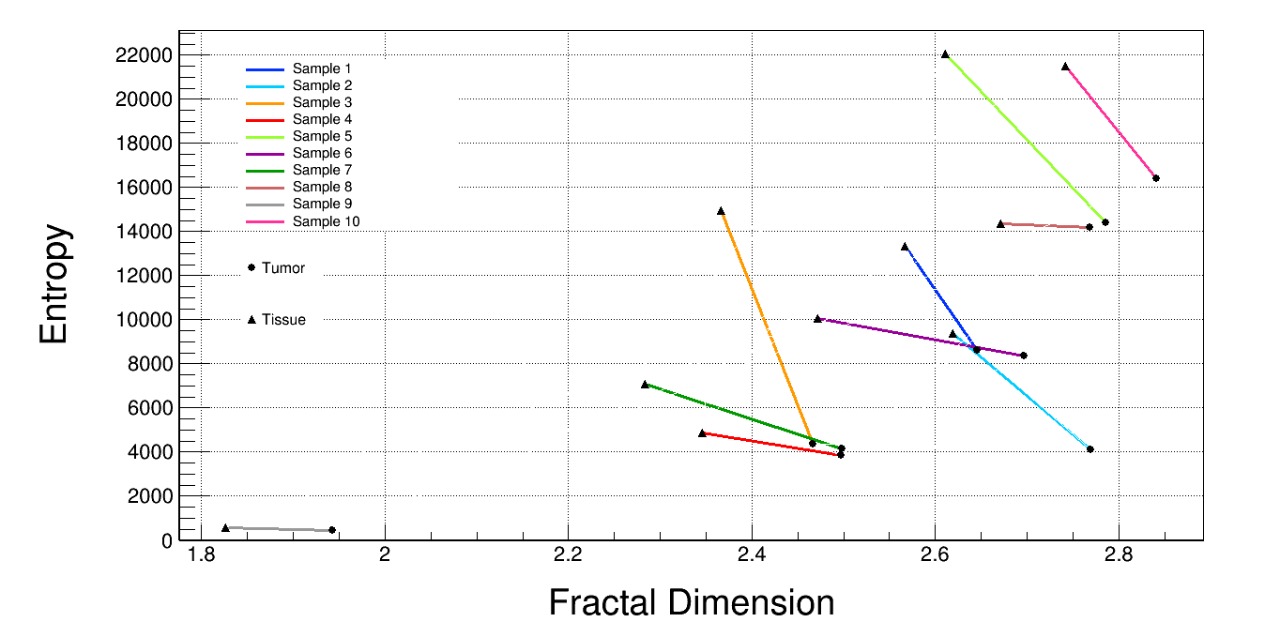}
\end{center}
\caption{Values of FD and S for T and t areas. It can be seen that the trend of these  values between the areas are the same for all samples. }
\label{T_analysis}
\end{figure}

The MI were given in a pixel array of 1024$\times$1024 in a PGM format, locating the origin in the lower left corner ~\cite{tumor_mammo}. The images were export to a PNG format,  an image in this format was required as an input data for \textit{TASImage Class Reference}. For this analysis, it was taken the malignant tumor MI, leaving a comparison between benign tumors for future analysis. As an example, in Figure~\ref{areas2} is shown one of the MI and its MIH. Diagnosis gives the coordinates of (338,314) for the center of T. The coordinates for the highest intensity value in the MIH is (-0.320313,-0.400391) or (346,310) element in the intensity array. It is simple to make the correspondence of 1024 size to the MIH axis size. To show consistency between the diagnosis and the measured value, in Table~\ref{compar} are shown the percentage  error for the coordinates $x$ and $y$, using 
\begin{equation}
    P_{error}i=\displaystyle\frac{|m_i-d_i|}{d_i}.
\end{equation}
Where, $m$ is the element value in the array, $d$ is the value giving by diagnostic and $i=x,y$ or rows and columns. Only ten samples are shown to avoid piling up of data, however, the same consistency was obtained for all samples. The following results are displayed for these ten samples.\\
\begin{center}
    \textit{Fractal dimension and entropy results}
\end{center}
The FD and S values for each area for the ten T-mammograms are shown in Figure~\ref{T_analysis}. From which, it can be seen that the FD has a higher value for T-area than b-area. While the S value is higher for b-area than T-area. This trend is the same for all samples, then, it was possible to distinguish between T and b areas from  the values of FD and S. The explanation of this trend is explained in Section~\ref{4}.

\begin{table}[htbp]
\caption{Percentage error between the diagnostic coordinate and the transformation coordinate.}
\label{compar}
\centering
\smallskip
\begin{tabular}{| c | c | c |}
\hline
Sample & $P_{error}x$(\%) & $P_{error}y$(\%) \\
\hline
1 & 2.36 & 5.75 \\
\hline
2 & 2.93 & 6.42\\
\hline
3 & 1.77  &  0.35\\
\hline
4 & 1.06 & 4.12\\
\hline
5 & 0.18 & 8.51\\
\hline
6 & 2.36 & 0.02\\
\hline
7 & 0.02 & 3.45\\
\hline
8 & 4.02 & 0.01\\
\hline
9 & 1.85 & 2.65\\
\hline
10 & 1.42 & 4.16\\
\hline
\end{tabular}
\end{table}

\begin{figure}[htbp]
\begin{center}
\includegraphics[width=0.4\textwidth]{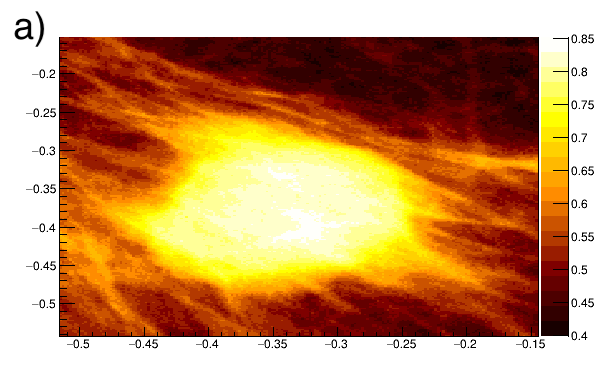}
\includegraphics[width=0.4\textwidth]{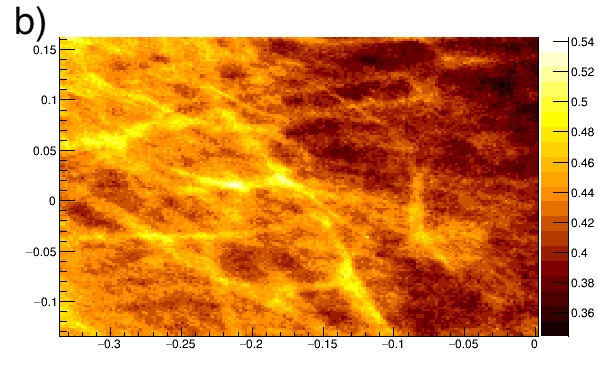}
\end{center}
\caption{a) T and b) t ROIs from Figure~\ref{areas2}, chosen according to diagnosis.}
\label{T_and_t}
\end{figure}

\begin{figure}[htbp]
\begin{center}
\includegraphics[width=0.4\textwidth]{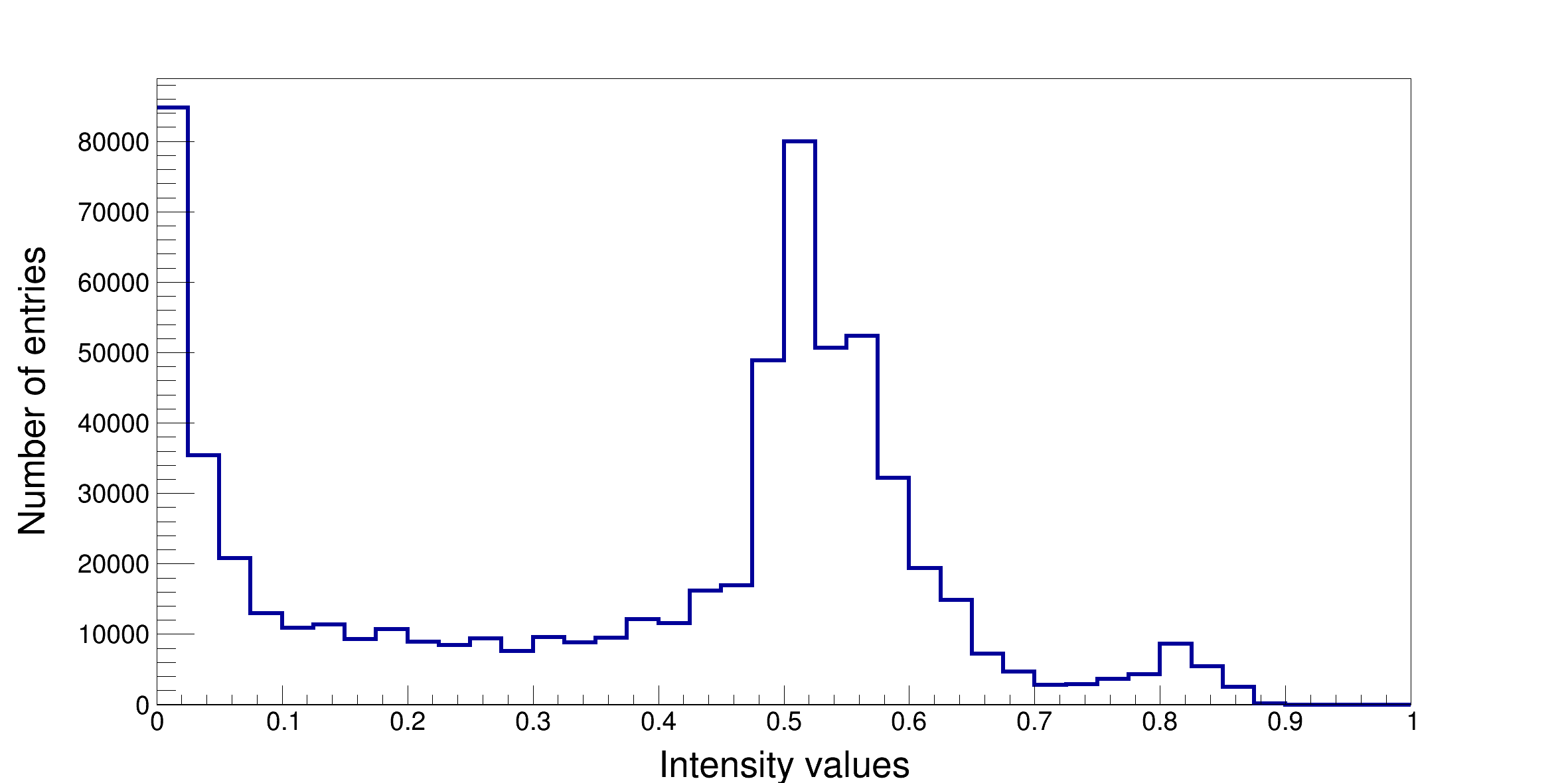}
\end{center}
\caption{Intensity distribution values from Figure~\ref{areas2}.}
\label{distribution}
\end{figure}

\begin{figure}[htbp]
\begin{center}
\includegraphics[width=0.4\textwidth]{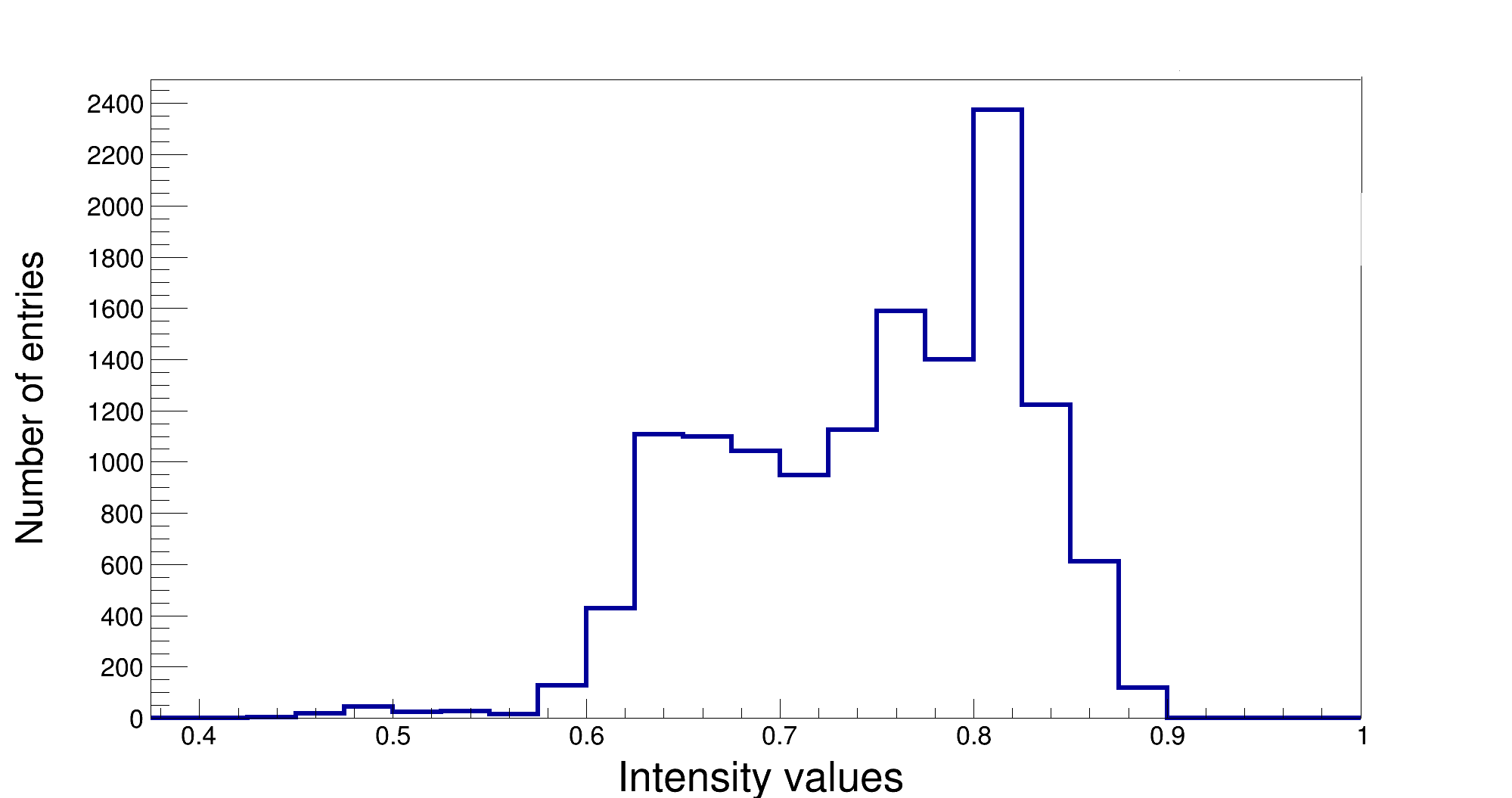}
\end{center}
\caption{Intensity T distribution values from Figure~\ref{T_and_t}}
\label{tumor_distribution}
\end{figure}

\begin{figure}[htbp]
\begin{center}
\includegraphics[width=0.4\textwidth]{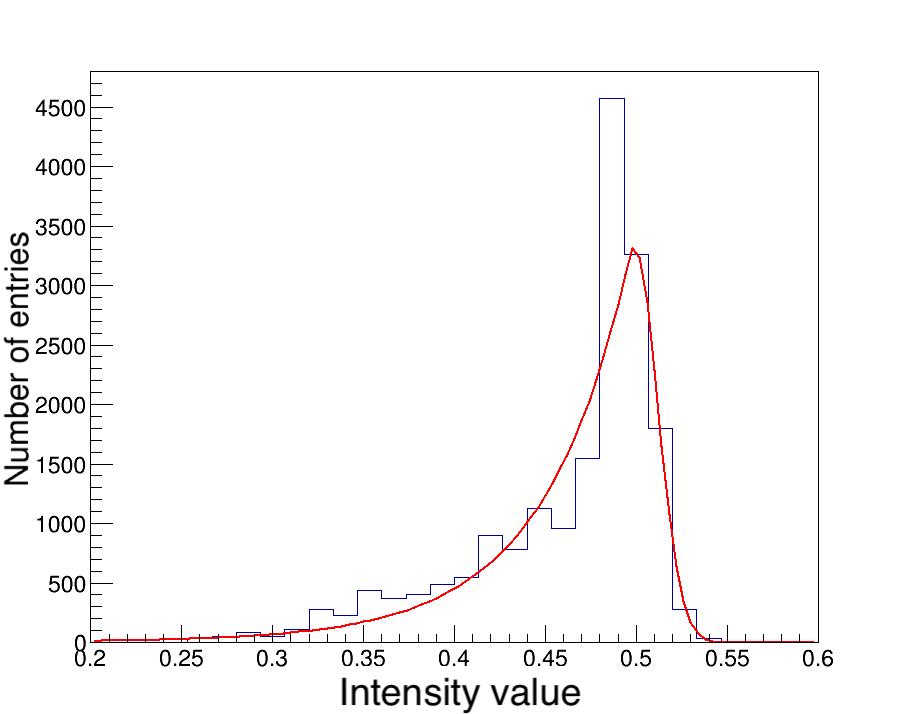}
\end{center}
\caption{Crystal Ball fit intensity distribution values from a b-area of Figure~\ref{T_and_t}. The values of $\sigma=0.0126\pm0.004$ and $mean=0.4990\pm0.0007$ from the fit were found.}
\label{tissue_distribution}
\end{figure}

\begin{center}
    \textit{Statistical analysis}
\end{center}
An intensity statistical analysis was proposed in order to show another method to distinguish between the areas. As an example, in Figure~\ref{T_and_t} are shown the T and b ROIs (MIH)  according to diagnosis for the MI of Figure~\ref{areas2}. To start with the analysis, in Figure~\ref{distribution} is shown the  intensity distribution of all sample. In   Figure~\ref{tumor_distribution} is shown the intensity distribution of T-area. Finally, in Figure~\ref{tissue_distribution} is shown the Crystal Ball fit  of the intensity distribution of a b-area selected, where were found the values of  $\sigma=0.0126\pm0.0040$ and $mean=0.4990\pm0.0007$.

Comparing Figure~\ref{distribution} and Figure~\ref{tumor_distribution}, it is simple to note that there are more intensity values from 0.5 to 0.6 in all distribution than in T-area. This means that in T-area there are intensity values that match with tissue. Once it has been characterized the b-area, as a first approximation, it was possible to separate T and b areas by choosing the maximum intensity value in T-area (being 0.890625) and the mean value in b-area.  Then, the maximum intensity value was located at 31$\sigma$ from the mean value in b-area. In Table~\ref{Info_Tumors} are shown the $\sigma$ and $mean$ values for the fit distribution in b-area and also it is shown the maximum value for the ten MI. Then, it is simple to note that the maximum intensity value (located in T-area) for each sample is bigger than 3$\sigma$ from the mean in their respective b-area,  i.e., the highest intensity value is located more than 99.6\% from the mean intensity values in b-area. Therefore, for a given intensity value greater than $mean+3\sigma$, it is located in T-area. This analysis was repeated for all MI and the same results were obtained. 
 
 \begin{table}[htb]
\centering
\begin{tabular}{|c|c|c|c|}
\hline
 & \multicolumn{2}{c|}{Fit values distribution in b-area} & Maximum intensity\\
\cline{2-3}
Sample & $mean$ & $\sigma$ & value in b-area\\
& & & \\
\hline
1  & 0.4317$\pm$0.0002 & 0.0224$\pm$0.0001 & 0.851565\\ \cline{1-4}
2  & 0.4923$\pm$0.0007 & 0.0126$\pm$0.0004 & 0.890625\\ \cline{1-4}
3  & 0.7898$\pm$0.0003 & 0.0241$\pm$0.0002 & 0.945313\\ \cline{1-4}
4  & 0.5117$\pm$0.0003 & 0.0222$\pm$0.0002 & 0.820313\\ \cline{1-4}
5  & 0.3226$\pm$0.0005 & 0.0169$\pm$0.0004 & 0.886719\\ \cline{1-4}
6  & 0.6299$\pm$0.0004 & 0.0245$\pm$0.0002 & 0.902344\\ \cline{1-4}
7  & 0.4813$\pm$0.0003 & 0.0269$\pm$0.0002 & 0.886719\\ \cline{1-4}
8  & 0.6565$\pm$0.0002 & 0.0259$\pm$0.0001 & 0.835938\\ \cline{1-4}
9  & 0.5894$\pm$0.0006 & 0.0189$\pm$0.0004 & 0.808594\\ \cline{1-4}
10 & 0.6342$\pm$0.0002 & 0.0240$\pm$0.0001 & 0.914063\\ \cline{1-4}
\end{tabular}
\caption{First column: Sample. Two next columns: $mean$ and $\sigma$ fit values from the fit intensity distribution in b-area. Last column: Maximum intensity value in the mammogram, i.e., the maximum value in b-area.}
\label{Info_Tumors}
\end{table}

\begin{figure}[htbp]
\begin{center}
\includegraphics[width=0.4\textwidth]{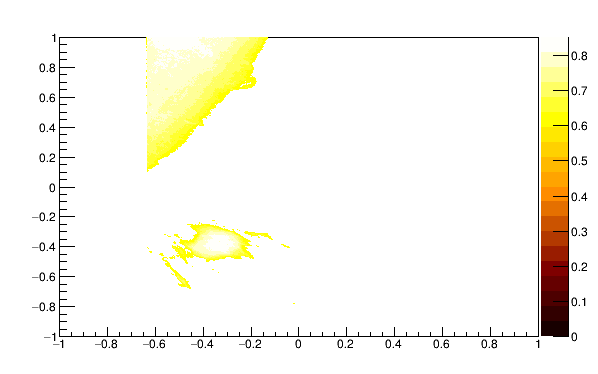}
\end{center}
\caption{Results of having selected intensity values greater than 0.6 in Figure~\ref{areas2}.}
\label{T_mammo_cuts}
\end{figure}

\begin{figure}[htbp]
\begin{center}
\includegraphics[width=0.4\textwidth]{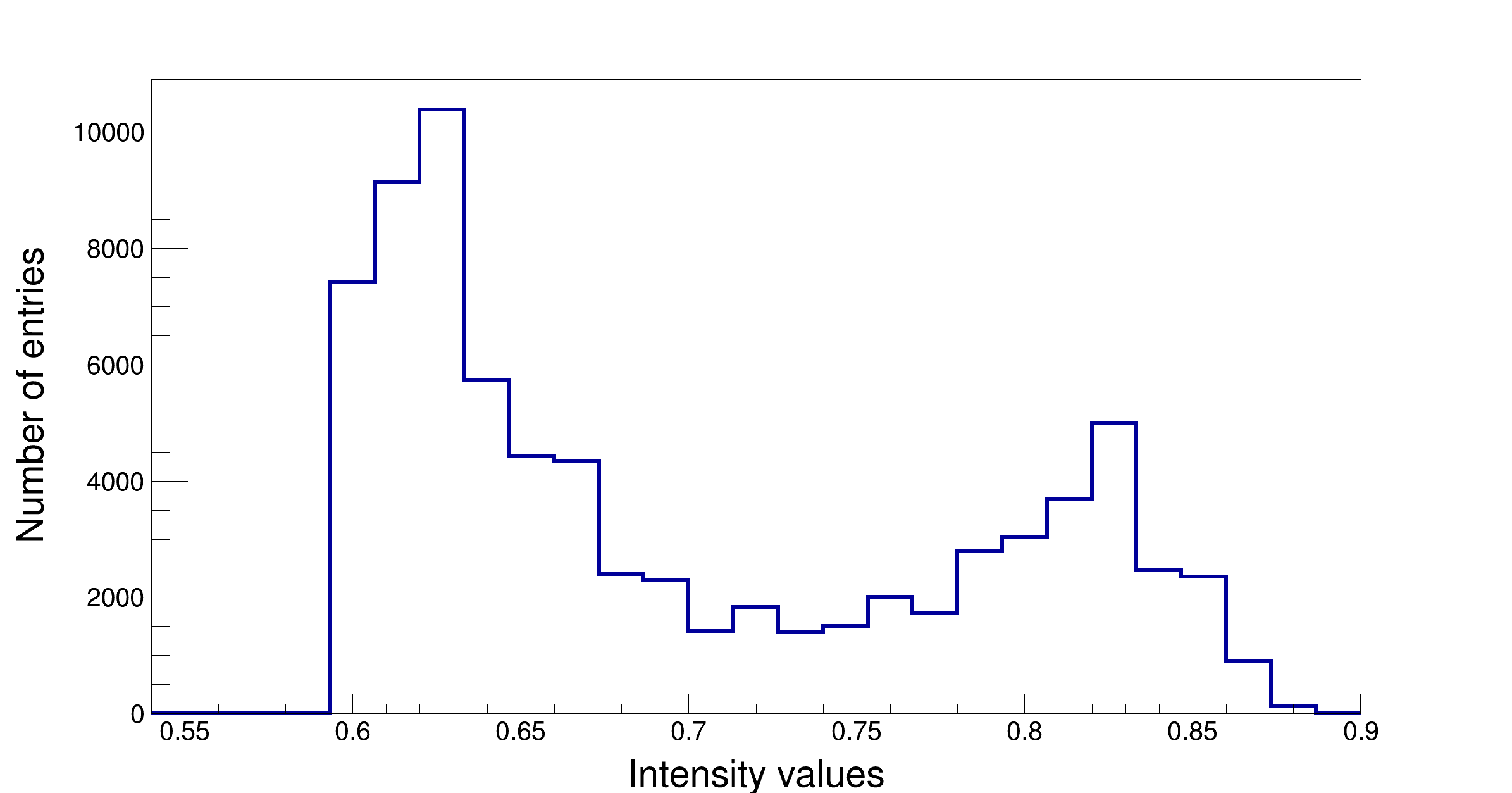}
\end{center}
\caption{Intensity values distribution after having selected intensity values greater than 0.6 in Figure~\ref{areas2}.}
\label{T_mammo_intensity_cuts}
\end{figure}

Another treatment was made in order to distinguish T and b areas. It consisted in taking the average intensity value of these areas, these values are shown in Table~\ref{T_average}. From this Table, it is possible to infer a cut in a threshold to exclude the section of b-area and distinguish  T-area. As an example, in Figure~\ref{T_mammo_cuts} is shown the intensity  values greater than 0.6 from Figure~\ref{areas2} to highlight T-area and the corresponding intensity distribution is shown in Figure~\ref{T_mammo_intensity_cuts}. It can be noticed other area on the top left, the diagnosis does not specify what it is, it may be a problem in the taking of the study, as already mentioned in the Introduction or a lot of tissue due to crushing of the breast, however, the maximum intensity value is located in T-area.

\begin{table}[htbp]
\caption{Average intensities values in T-mammograms for t and T areas.}
\label{T_average}
\centering
\smallskip
\begin{tabular}{| c | c | c |}
\hline
Sample & b-area & b-area \\
\hline
1 & 0.629 & 0.851\\
\hline
2 & 0.855 & 0.917\\
\hline
3 & 0.632 &0.945 \\
\hline
4 & 0.707 & 0.941\\
\hline
5 & 0.652 & 0.945\\
\hline
6 & 0.640 & 0.820\\
\hline
7 & 0.882 & 0.894\\
\hline
8 & 0.718 & 0.808\\
\hline
9 & 0.765 & 0.878\\
\hline
10 & 0.820 & 0.890\\
\hline
\end{tabular}
\end{table}

\begin{figure}[htbp]
\begin{center}
\includegraphics[width=0.4\textwidth]{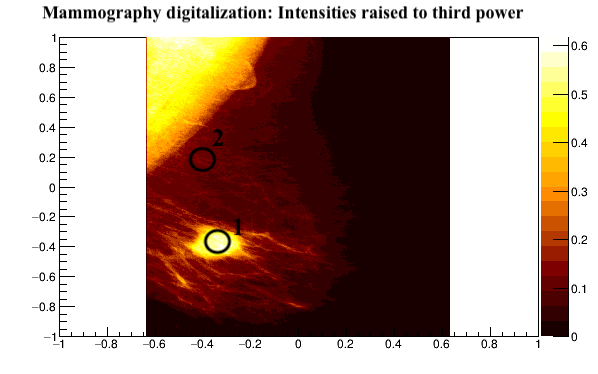}
\includegraphics[width=0.4\textwidth]{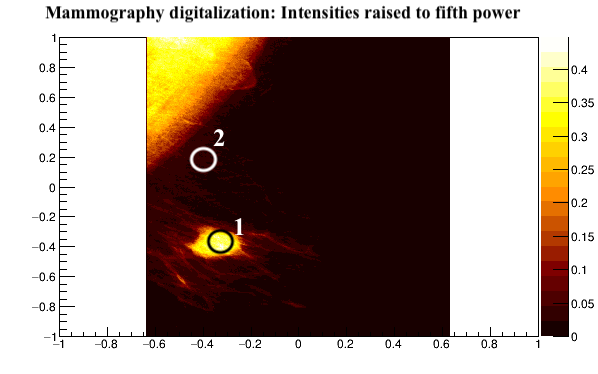}
\includegraphics[width=0.4\textwidth]{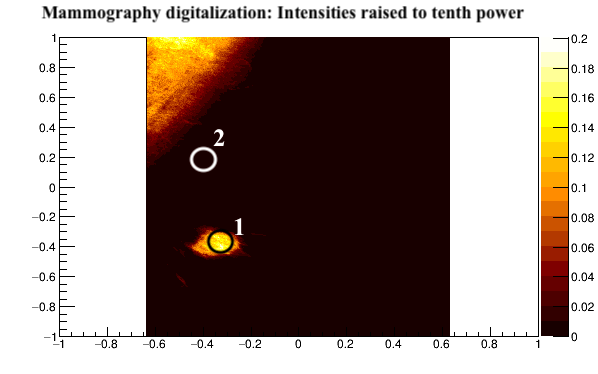}
\end{center}
\caption{Results of having raised to third,  fifth and tenth power the image of Figure~\ref{areas2}. Comparing the images, it is simple to see that  b-area becomes prominent.}
\label{T_power}
\end{figure}

\begin{center}
    \textit{Highlight tumor values by powers}
\end{center}
Finally, when raising to certain powers the intensity values of MIH, the area of the tumor stands out. As an example, in Figure~\ref{T_power} are the results by raising the intensity values from Figure~\ref{areas2} to third, fifth and tenth power. The T-area becomes prominent.

\subsection{MC-mammograms analysis}
\begin{figure}[htbp]
\begin{center}
\includegraphics[width=0.5\textwidth]{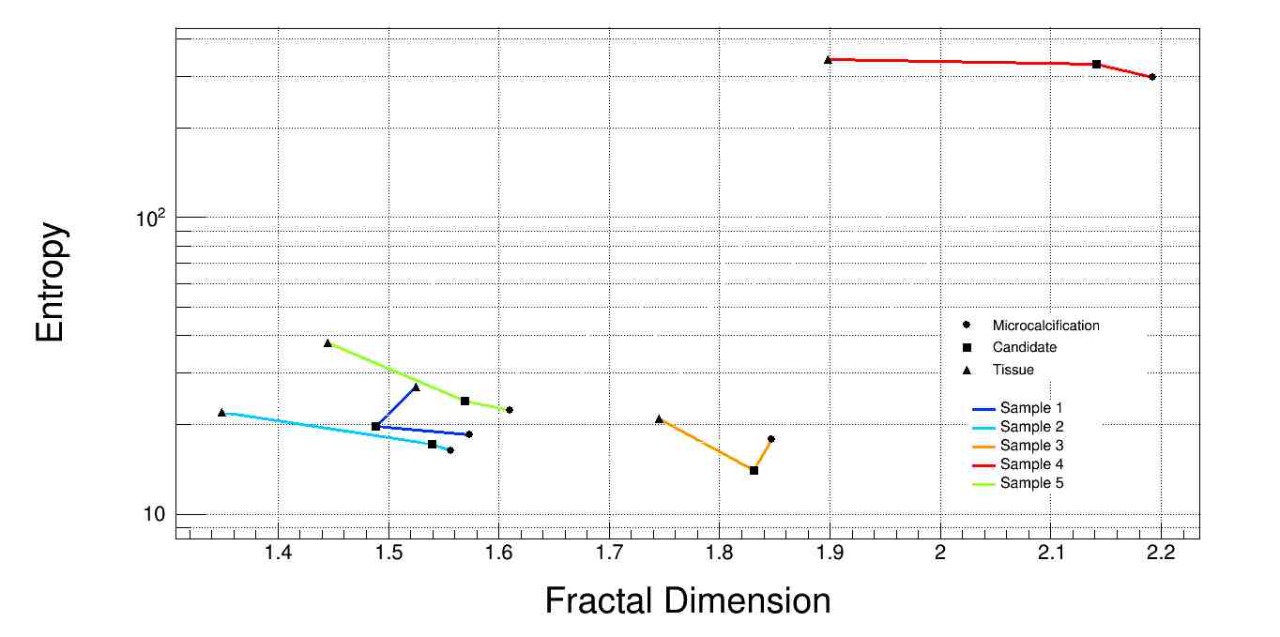}
\end{center}
\caption{Correlation between FD and S values for the MC, C and t areas from  MC-mammograms.}
\label{MC_analysis}
\end{figure}

For these MI, it highlights the C-area, its central pixel always was found to have the second highest intensity value after MC. These images were provided in a PNG format.  As an example, the coordinates for the MC from Figure~\ref{sigmapmt} are (0.681957, 0.622490). These MI,  cannot be compared with coordinates in the MHI (as in the T-mammograms), because the diagnosis was made by the diagnosis given  by Dra. Herrera. However, the location of the MC matches the diagnosis.\\
\begin{center}
    \textit{Fractal dimension and entropy analysis}
\end{center}
The FD and S values for the five MC-images are shown in Figure~\ref{MC_analysis}. From which, it can be seen that the FD has a higher value for MC than the C and b values. While the S value is higher for b-area than the MC-area. The FD and S values of C can fluctuate, because there are more intensity values compared to MC and b areas, however its FD value is always less than the MC value. Then, the MC-area can be distinguished. A more precise discussion is shown in Section~\ref{4}.

\begin{figure}[htbp]
\begin{center}
\includegraphics[width=0.4\textwidth]{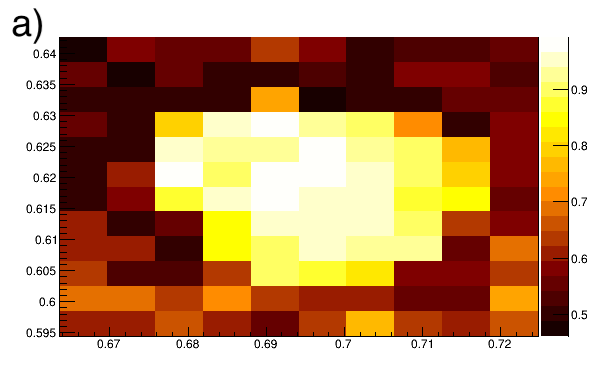}
\includegraphics[width=0.4\textwidth]{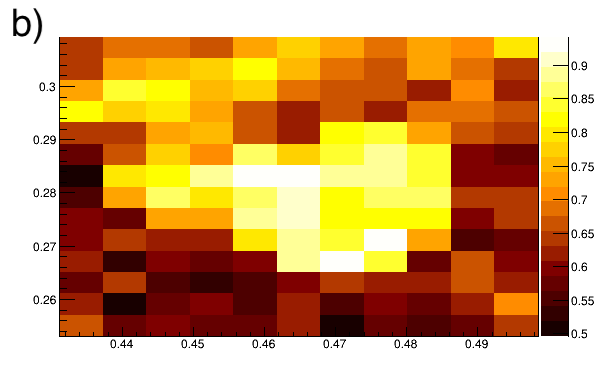}
\includegraphics[width=0.4\textwidth]{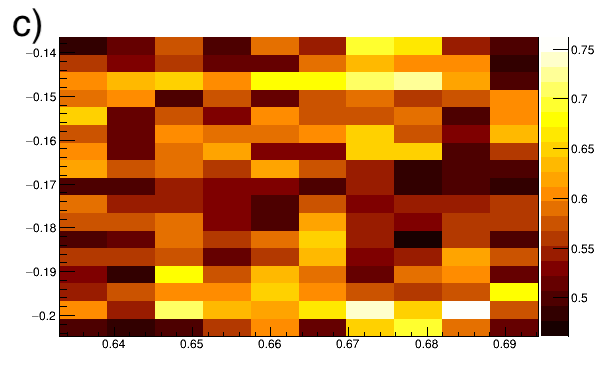}
\end{center}
\caption{a) MC, b) C and c) t area choosing to enclose MC-area from Figure~\ref{sigmapmt}. It can be seen the low statistic compare it to Figure~\ref{T_and_t}.}
\label{MC_and_C_t}
\end{figure}

\begin{figure}[htbp]
\begin{center}
\includegraphics[width=0.5\textwidth]{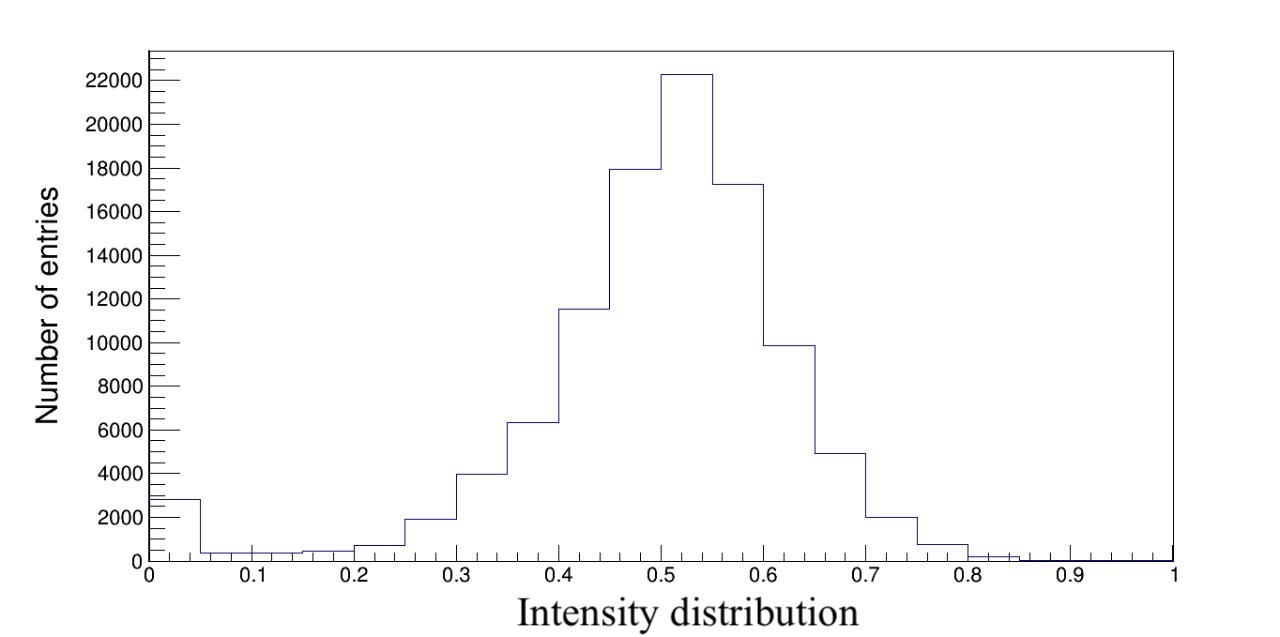}
\end{center}
\caption{Intensity values distribution for Figure~\ref{sigmapmt}.}
\label{MC_all}
\end{figure}

\begin{figure}[htbp]
\begin{center}
\includegraphics[width=0.5\textwidth]{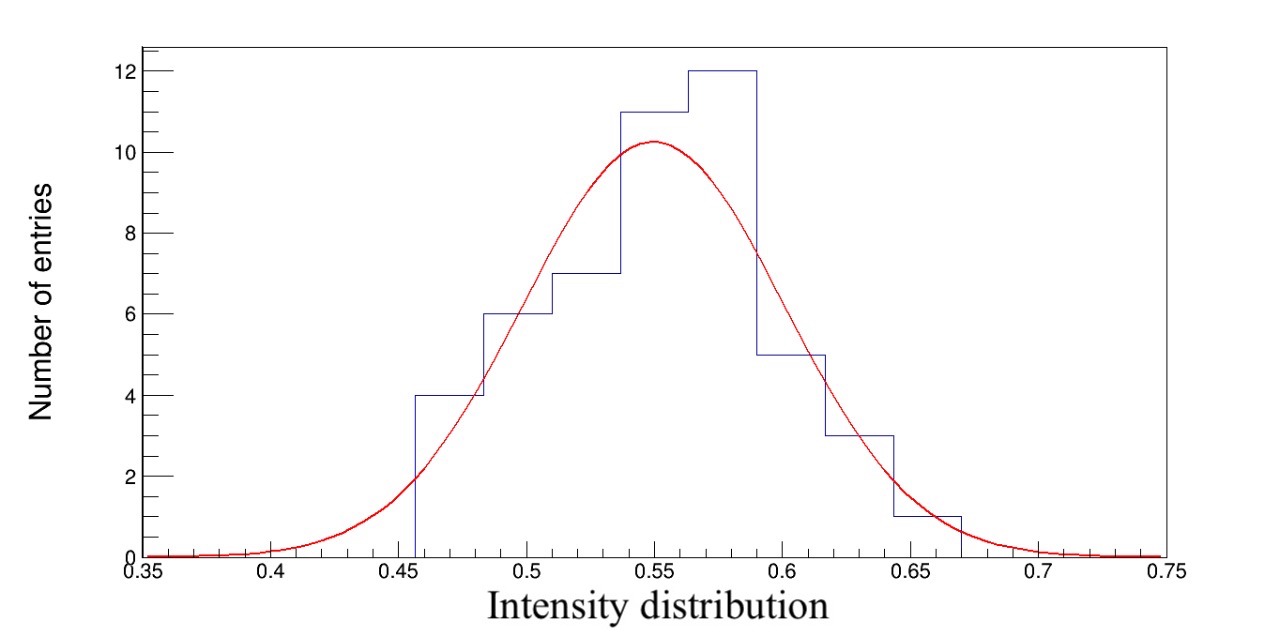}
\end{center}
\caption{Intensity t values distribution by choosing one of many b-areas from Figure~\ref{sigmapmt}. The Gaussian fit values are: $mean=0.5496\pm0.0093$ and $\sigma=0.0510\pm0.0087$.}
\label{MC_t}
\end{figure}

\begin{center}
    \textit{Statistical analysis}
\end{center}
Similar to T-mammograms, an intensity value statistical analysis was made. The statistic in each MIH was lower than in  the previous case, however, it was enough for the analysis in  b-area. As an example, in Figure~\ref{MC_and_C_t} are shown MC, C and b areas from Figure~\ref{sigmapmt}, the sizes of this areas were choosing to enclose first the MC-area. In Figure~\ref{MC_all} is shown the intensity distribution of all sample. Note that in the b-area there are many more values than in MC-area. The values distribution, of one of many b-areas, are shown in Figure~\ref{MC_t}, the low statistic was due to the size of the area selected, regarding the size of MC-area. The fit distribution  values are  $mean=0.5496\pm0.0093$,  $\sigma=0.0510\pm0.0087$ and $\chi^2/ndof=2.46$. According to the MIH, the maximum intensity value of all image (central MC intensity) was 0.992188. Then, the MC maximum intensity value was located more than 3$\sigma$ from the $mean$ b-value.

This analysis was made for all samples of MC MI and in Table~\ref{Info_MC} are shown the fit values of b-area intensity distribution, the maxim intensity value (MC central value) and the maximum intensity value of C-area.

 \begin{table}[htb]

\centering

\begin{turn}{90}
\begin{tabular}{|c|c|c|c|c|}
\hline
 & \multicolumn{2}{c|}{Fit values distribution in b-area} & Maximum intensity & Maximum intensity\\
\cline{2-3}
Sample & $mean$ & $\sigma$ & value in C-area & value in MC-area\\
& & & & \\
\hline
1  & 0.613$\pm$0.01390 & 0.0636$\pm$0.01211 & 0.898438 & 0.996094\\ \cline{1-5}
2  & 0.550$\pm$0.00933 & 0.0510$\pm$0.00868 & 0.957031 & 0.964844\\ \cline{1-5}
3  & 0.479$\pm$0.00835 & 0.0511$\pm$0.00706 & 0.929688 & 0.941406\\ \cline{1-5}
4  & 0.189$\pm$0.00131 & 0.0329$\pm$0.00109 & 0.902344 & 0.988281\\ \cline{1-5}
5  & 0.558$\pm$0.00933 & 0.0510$\pm$0.00807 & 0.941406 & 0.992188\\ \cline{1-5}
\end{tabular}
\end{turn}

\caption{First column: Sample. Two next columns: $mean$ and $\sigma$ fit values from the fit intensity distribution in b-area. Last columns: Maximum intensity values in C-area and MC-area.}
\label{Info_MC}
\end{table}

\begin{figure}[htbp]
\begin{center}
\includegraphics[width=0.4\textwidth]{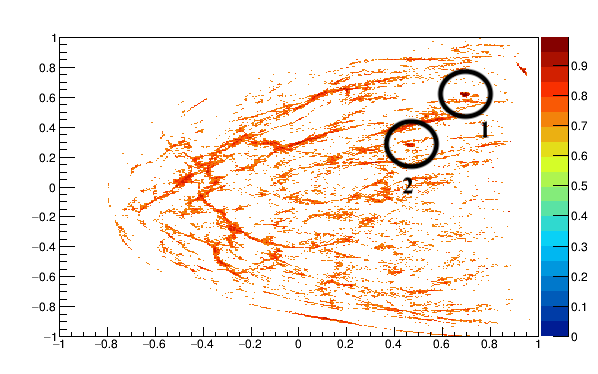}
\end{center}
\caption{Results of having selected intensity values greater than 0.7 from Figure~\ref{sigmapmt}}
\label{MC_cut}
\end{figure}

As it was mentioned, the statistic in C and MC areas was very small, less than 50 entries. The main goal was to distinguish the b and MC areas. Then, from Table~\ref{Info_MC} it is simple to note that the MC values are located more than 3$\sigma$ from the mean value in b-area. Then, MC can be distinguished.\\
As it was made for T-mammograms, the average values were calculated for each area to infer the cuts in a threshold and distinguish the MC-area. In Table~\ref{MC_average} are shown the average values of the MC and b areas. As an example, taking values greater than 0.7, the Figure~\ref{sigmapmt} was transformed as the Figure~\ref{MC_cut} shown. Clearly, the choosing of a bigger cut, the MC-area will be more prominent.

\begin{table}[htbp]
\caption{Average intensities values in MC-mammograms for t and MC-areas.}
\label{MC_average}
\centering
\smallskip
\begin{tabular}{| c | c | c |}
\hline
Sample & b-area & MC-area\\
\hline
1 & 0.574 & 0.902\\
\hline 
2 & 0.495 & 0.916\\
\hline
3 & 0.529 & 0.965\\
\hline
4 & 0.464 & 0.930\\
\hline
5 & 0.242 & 0.811\\
\hline
\end{tabular}
\end{table}

\begin{figure}[htbp]
\begin{center}
\includegraphics[width=0.4\textwidth]{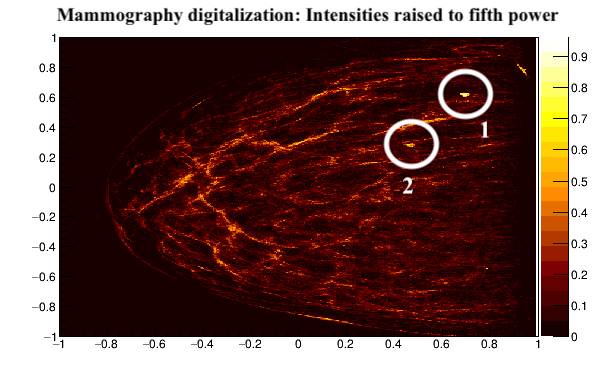}
\includegraphics[width=0.4\textwidth]{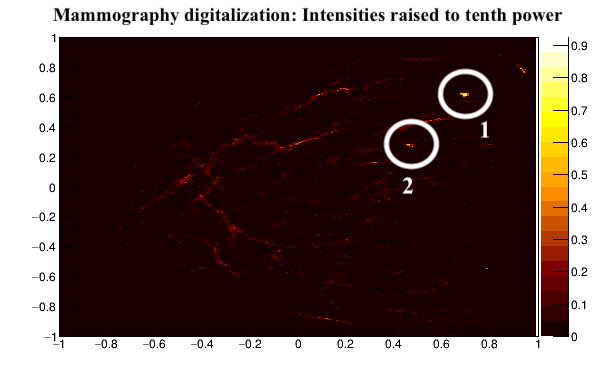}
\includegraphics[width=0.4\textwidth]{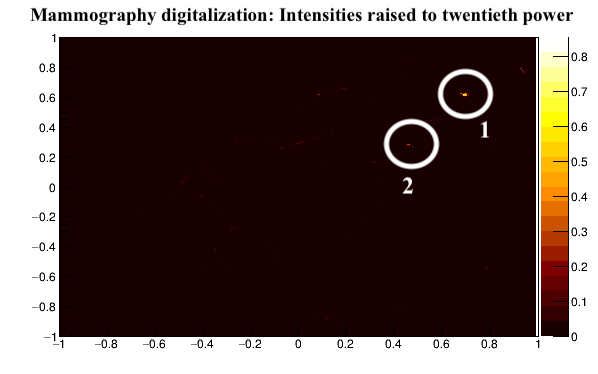}
\end{center}
\caption{Results of having raised to a) third, b) sixth and c) fifteenth power the image of Figure~\ref{sigmapmt}. A white spot is protruding, which corresponds to the MC.}
\label{power}
\end{figure}

\begin{center}
    \textit{Highlight microcalsification values by powers}
\end{center}
Finally, in Figure~\ref{power}, it is shown the raised intensity values to third, sixth and fifteenth power from Figure~\ref{sigmapmt}. Then, it can be seen that highlights a white spot, which, corresponds to the MC location. Same results were obtained for all MI.

\section{Discussion and conclusions}\label{4}
When transforming a mammogram image into a 2D histogram, where two axis are the location coordinate system for the pixels and the palette  is for their intensity  value, the most important result was that the highest intensity pixel value in a MIH was located in the T/MC-area of the MI. For  the case of dense-glandular mammograms, the location of the malignant tumor can be complicated, then, this method can help these cases. It is shown a simple method for the localization T/MC-area. From Figures~\ref{T_analysis} and \ref{MC_analysis} it was shown that there is a correlation between FD and S for b-area and T/MC-area. Note that the trend is the same, independent if it is a T or a MC. It can conclude that the value of FD is higher for T/MC-area than the corresponding b-area, i.e, when selecting two areas in a mammogram, it is possible to distinguish them according to their intensity values. This result can be understood by the following: the size of the pixel in each mammogram is constant, then, according with Eq.~\ref{fraceq}, the FD is proportional to $A(\epsilon,i_\epsilon)$ which is proportional to intensity values in that area, thus, as the pixel with highest value is located in the T/MC-area, the FD value is higher for this area. The higher value of S is like in thermodynamics, which usually describe the system's disorder, in this case, the uniformity or non-uniformity of the intensity values.  Therefore, the conclusion is that in the b-area there is a great variety of intensity values, while in T/MC-area the intensity values are uniform. Finally, the different values between FD and S for each sample may be due to fatty tissue in each breast, the brightness by which the image was obtained, etc.\\
To have another discrimination method, it was  made a statistical intensity analysis to characterize b-area and then exclude it to finally keep the intensity values in the T/MC-area. From Tables~\ref{Info_Tumors} and \ref{Info_MC} it was possible to identify b and T/MC areas, i.e., a second method was shown, in which it was possible to distinguish a T/MC from a mammogram, since the intensity  values were located more than 3$\sigma$ from the mean in b-value. From Tables~\ref{T_average} and \ref{MC_average}, it was suggested to make a cut in a threshold up to 0.7 in intensity value to keep the T/MC-area, because for this analysis, it was not important to know the shape of T/MC. The results of this analysis can help to the diagnostic, to locate the T/MC, even when the breast is too dense, for which, it cannot be distinguished by naked-eye. Also, it was shown that with the MIH, it was simple to highlight the T/MC-area, by raising the intensity values to some power.\\
Considering both sets of MI, it is simple to note that choosing  a threshold greater than 0.6, the T and MC areas will stand out on the MIH or MI. Part of this analysis was made for malignant tumors, then a future work can be made for benignat tumors and find whether or not if there is a difference from this results.\\
Finally, it is important to note that the image format of the original mammography is not as relevant when using this method, due to the transformation to the MIH: an image can be transformed into an information pixel intensity.

\section*{Acknowledgments}
We thank would like to thank Dr. Karla Herrera for her support providing all the Microcalcification mammograms. We would also like to thank the Digital Mammography Home Page of the University of South Florida, for access to the Tumor mammograms. Both sets of data were provided with their corresponding clinical evaluations, information without which this work would have never been carried out.

\begin{minipage}{.5\textwidth}
\centering
\bf \large Refrerence
\end{minipage}

\end{document}